# HIGH-Q OPTICAL NANOCAVITIES IN BULK SINGLE-CRYSTAL DIAMOND

*Michael J. Burek[a], Yiwen Chu[b], Madelaine S.Z. Liddy[c], Parth Patel[c], Jake Rochman[c], Srujan Meesala[a], Wooyoung Hong[a,d], Qimin Quan[a,d], Mikhail D. Lukin[b], and Marko Lončar[a,†]*

[a.] School of Engineering and Applied Sciences, Harvard University, 29 Oxford Street, Cambridge, MA 02138, USA

[b.] Department of Physics, Harvard University, 17 Oxford Street, Cambridge, MA 02138, USA

[c.] University of Waterloo, 200 University Avenue West, Waterloo, ON N2L 3G1, Canada

[d.] Rowland Institute at Harvard, Harvard University, 100 Edwin H. Land Blvd., Cambridge, MA 02142, USA

[†] Corresponding author contact: E-mail: loncar@seas.harvard.edu. Tel: (617) 495-579. Fax: (617) 496-6404.

**Single-crystal diamond, with its unique optical, mechanical and thermal properties, has emerged as a promising material with applications in classical and quantum optics. However, the lack of heteroepitaxial growth and scalable fabrication techniques remain major limiting factors preventing more wide-spread development and application of diamond photonics. In this work, we overcome this difficulty by adapting *angled-etching* techniques, previously developed for realization of diamond nanomechanical resonators, to fabricate racetrack resonators and photonic crystal cavities in bulk single-crystal diamond. Our devices feature large optical quality factors, in excess of $10^5$, and operate over a wide wavelength range, spanning visible and telecom. These newly developed high-Q diamond optical nanocavities open the door for a wealth of applications, ranging from nonlinear optics and chemical sensing, to quantum information processing and cavity optomechanics.**

Contemporary integrated nanophotonic platforms all have one important feature in common: they consist of a device layer – an optically thin film – supported by a substrate of a *different* material. The substrate provides optical isolation either by having a smaller refractive index than the device layer, or by its selective removal[1]. Single-crystal diamond is one example from an extensive list of materials – many with attractive optical properties – for which *high quality* thin film heterolayer structures do not exist. Despite wafer-scale polycrystalline diamond thin films on foreign substrates being readily available, these films typically exhibit inferior properties due to scattering and absorption losses at grain boundaries, significant surface roughness, and large interfacial stresses[2-4]. In the absence of suitable heteroepitaxial diamond growth, substantial efforts by the diamond photonics and quantum optics community have focused on novel processing techniques to yield nanoscale single-crystal diamond optical elements[5-11]. For the most part, these efforts have involved heterogeneous integration of single-crystal diamond slabs (~ 5 to 30 μm thick) onto supporting silica substrates, with subsequent oxygen plasma etching to thin the slab near a target thickness ~ 500 nm or less. Ring resonators[12-15] and photonic crystal cavities[16-17] have been realized in such thinned diamond membranes, with recent results[18] demonstrating ultra-high quality factors (Q) in excess of $10^6$. While this approach remains promising[19], complications due to material handling, scalability, repeatability, and sheer difficulty of removing tens of microns of diamond while preserving uniform hundred-nanometer scale films limit this approach significantly. In this work we demonstrate state-of-the art nanophotonic resonators in single-crystal diamond substrates, realized by *angled-etching*[20]: an unconventional, yet scalable fabrication technique. Our approach, depicted in Figures 1 (a) and (b) with details summarized in *Methods*, employs anisotropic oxygen-based plasma etching at an oblique angle to the substrate surface, resulting in suspended structures with triangular cross-section. Angled-etching is performed in a standard inductively coupled plasma-reactive ion etcher (ICP-RIE), however the diamond substrate is housed within a specifically designed aluminium Faraday cage which modifies the trajectory of the incident plasma ions towards the sample surface (refer to *Supplementary Information* for additional description of the angled-etching technique and Faraday cage designs). Importantly, diamond resonators fabricated in this way, shown in Figure 1 (c) and (d), feature optical Q-factors on par with those found in devices realized in standard materials (e.g. silicon), using conventional (planar) microfabrication techniques.

Of the myriad of on-chip optical cavities demonstrated to date, ring and racetrack resonators are arguably the most ubiquitous[21]. Conceptually, the optical cavity is a waveguide looped back on itself, and the resonance is formed when the optical path length is an integer multiple of the wavelength. In the context of angled-etching, creating a free-standing looped waveguide represents a challenge, since suspended devices must be supported by at least one physical attachment to the bulk substrate. While free-standing wheel-and-spoke optical cavity structures[20,22-23] are an intuitively obvious solution, spoke attachment points to the looped triangular cross-section waveguide are difficult to fabricate, resulting in significant scattering losses. To circumvent this, we have developed novel vertical support structures, shown in Figure 2 (a) - (c), and used it to realize single-crystal diamond racetrack resonators. This was accomplished by positively tapering the width of the 20 μm long straight portions of the racetrack resonator by 15 % of the nominal value. Note, a ~ 50° etch angle was used to fabricate diamond racetrack resonators shown in Figure 2 (see *Supplementary Information* for cross-sectional view). In angled-etching[20], wider features require more time to be fully released from the substrate, since the structure thickness ($t$) is intrinsically linked to its width ($w$) by the etch angle ($\theta$), via the relationship $t = w/(2\tan\theta)$. As a consequence, wider sections can remain attached to the substrate resulting in a pedestal-like support at their center. At the same time, the tapered nature of the vertical support structure minimizes optical losses (see *Supplementary Information*).

A typical normalized transmission spectrum collected by tunable laser and photodiode from diamond racetrack resonator is shown in Figure 2 (f). Fiber taper coupling[24] was used in the characterization of these devices (see *Methods*). Two distinct sets of transmission dips are observed, corresponding to fundamental TE-like and TM-like transverse modes of the structure, shown Figure 2 (d) and (e), respectively. Nearly critically coupled resonances displayed in Figure 2 (g) and (h) reveal loaded Q-factors of $Q_{L,TE}$ ~151,000 and $Q_{L,TM}$ ~ 113,000, where the subscript indicates the cavity mode transverse polarization. We note that the term loaded Q-factor refers to the Q-factor that includes losses due to fiber taper coupling, and at critical coupling is half the intrinsic Q-factor (i.e., $Q_L = \frac{1}{2}Q_i$). The latter is determined by losses due to scattering, material absorption, leakage to the substrate (if any), and waveguide bends/overlaps. From our measurements, we estimate intrinsic Q-factors to be $Q_{i,TE}$ ~ 302,000 and $Q_{i,TM}$ ~ 226,000. From measured Q-factors, an upper limit on the diamond waveguide transmission loss ($\alpha$) is estimated to be ~ 1.5 dB/cm for both guided modes via the relationship[25]: $\alpha \approx 2\pi n_g / Q_i \lambda$, where $n_g$ is the mode group index, and $\lambda$ is the resonant wavelength. While this loss value is roughly five times greater than that recently reported for single-crystal diamond waveguides fabricated via the membrane thinning approach[18], it is also an order of magnitude smaller than losses of polycrystalline diamond ring resonators[26].

We have also fabricated diamond photonic crystal nanobeam cavities[27] using our angled-etching approach. These devices consist of a waveguide perforated with a chirped lattice of elliptically-shaped air holes, which has been engineered to support resonances with ultra high Q-factors and ultra small mode volumes[28]. Figures 3 (a) - (c) display a representative single-crystal diamond nanobeam cavity fabricated (etch angle $\theta \sim 35°$, see *Supplementary Information* for view of cross-section) for operation in the telecom band. Since the nanobeam thickness and width are linked through angled-etching, global scaling of the nanobeam cavity dimensions results in tuning of the cavity resonance while maintaining all cavity figures of merit (i.e. Q-factor and mode volume). Therefore, the nanobeam cavity design used in this work is parameterized by the target fundamental TE-like cavity mode resonance wavelength, $\lambda_{TE}$. Our design[29] has the following parameters: a nanobeam width $w = 0.58\lambda_{TE}$, lattice constant (hole spacing) $a = 0.319\lambda_{TE}$, and elliptical hole minor radius $r = 0.087\lambda_{TE}$. Furthermore, to minimize the scattering and maximize the cavity Q, the major radius of the elliptical hole array is decreased quadratically, over 30 periods, from $r_1 = 0.145\lambda_{TE}$ at the center of the cavity, to $r_{30} = 0.087\lambda_{TE}$ at its end. We modeled the devices using finite-difference time-domain (FDTD) methods, and found they support both TM-like and TE-like resonances (with the fundamental TM-like resonance, $\lambda_{TM}$, located at $0.9\lambda_{TE}$), with representative mode profiles shown in Figure 3 (d) and (e) respectively. The dual mode nature of the triangular cross-section nanobeam cavities is of interest for applications in nonlinear optics and wavelength conversion[30-32]. Theoretical figures of merit for the fundamental cavity modes are Q-factors of $Q_{TM} \sim 1.3 \times 10^5$ and $Q_{TE} \sim 3.0 \times 10^6$, with mode volumes $V_{TM} \sim 2.55(\lambda/n)^3$ and $V_{TE} \sim 2.26(\lambda/n)^3$, (the subscript again refer to the cavity mode transverse polarization). Additionally, due to the gradual nature of the chirped lattice of air holes, the devices also support higher order longitudinal modes of both polarization. We note that devices shown in Figure 3 are based on a design with fundamental cavity resonances located at $\lambda_{TE} = 1680$ nm and $\lambda_{TM} = 1507$ nm respectively.

A normalized transmission spectrum of a representative diamond nanobeam cavity is shown in Figure 3 (f). Two sets of transmission dips are observed, and are attributed to cavity resonances: dips located near 1610 nm correspond to TE-like modes, while those located near 1490 nm correspond to TM-like modes. The experimentally obtained fundamental resonance wavelengths for the two modes indicate $\lambda_{TM,exp} \sim 0.92\lambda_{TE,exp}$, which is in good agreement with FDTD predictions. The absolute values of cavity resonances are blue shifted by roughly 5% from the target values, which is likely due to the uncertainty in the actual etch angle. The latter was previously estimated to deviate up to 2° degrees from the nominal value[33] (see *Supplementary Information* for additional cross-sectional analysis). High resolution spectra of fundamental TM-like and TE-like cavity modes are shown in Figure 3 (g) and (h) respectively. The loaded Q-factor of the fundamental TE-like cavity mode is remarkably high at $Q_{TE} \sim 183,000$ and compares very well to the state-of-the-art silicon photonic crystal nanobeam cavities realized by standard fabrication techniques. We note that the loaded Q-factor of the fundamental TM-like nanobeam cavity is approximately $Q_{TM} \sim 24,000$. Nearly an order of magnitude reduction in cavity Q for this mode is likely due to its localization at the bottom apex of the nanobeam, which increases its losses by scattering from overlap with etched surfaces and leakage into the diamond substrate.

Finally, to utilize the broadband nature of diamond, we explored the potential of our angled-etching approach to realize optical cavities operating in visible and near-IR. Visible diamond cavities are of great interest for the enhancement of emission properties of diamond's luminescent defects, such as the negatively charged silicon vacancy center (SiV⁻, zero phonon line at λ ~ 737 nm)[34-36] and, in particular, the negatively charged nitrogen vacancy center (NV⁻, with zero phonon line at λ ~ 637 nm and phonon side band up to nearly 800 nm)[37-39]. To realize visible band optical cavities in diamond, we scaled down all design parameters by a factor of approximately 2.5, and no additional modeling was needed. This design flexibility is an inherent property of angled-etching in which device thickness is coupled to its width. The same is not true for the planar technologies where one dimension is always fixed by the thickness of the device layer (e.g. a 220 nm thick silicon device layer in the case of silicon-on-insulator). Therefore, angled-etching allows for the integration of devices operating over a wide wavelength range (UV to mid-IR) to be easily integrated on the same diamond chip. Figure 1 (c) shows a fabricated visible band diamond racetrack resonators, with a ~ 500 nm wide suspended waveguide and 17.5 μm bend radius. In such devices, the material segment supporting the free-standing waveguide is estimated to be ~ 90 nm thick. A broadband normalized transmission spectrum, shown in Figure 4 (a), is obtained using a combination of tunable red laser (635 nm to 639 nm) and super-continuum source (see *Methods* for details). The insets of Figure 4 (a) display the fiber taper coupling position with the laser tuned off and on resonance; the diamond race track resonator lights up when the laser is resonant with the optical cavity. We note that due to the small coupling gap necessary at visible wavelengths, van der Waals attraction between the fiber taper and diamond device forced these measurements to be taken with the fiber taper touching the device. This ultimately limited coupling efficiency and measured cavity Q-factors. High resolution spectra, collected with the tunable laser, of racetrack resonator cavity modes located at approximately ~ 637 nm, are shown in Figure 4 (b). The measured loaded Q-factors of the cavity modes were 33,000 and 59,000. For cavity modes at longer wavelengths, accurate measurement of their Q-factors by a tunable laser was not possible. However, Q-factors for resonances near 800 nm estimated from spectra collected by super-continuum excitation exceed the resolution limit of the spectrometer, and thus were at least $10^4$. Therefore, free-standing diamond waveguides fabricated by angled-etching operate with low loss over a large, nearly 200 nm wide bandwidth that covers visible and near-infrared wavelengths.

Figure 1 (d) shows a representative diamond nanobeam cavity fabricated using the same design as previously described, with a target resonance for the fundamental TE-like cavity mode of $\lambda_{TE}$ = 710 nm. In order to characterize such structures, a free-space coupling technique was used in lieu of fiber taper coupling, given the challenge of obtaining proper fiber alignment to small visible nanobeam cavities. The free-space measurement set up (see *Methods*) allowed for in- and out-coupling of light at opposite ends of the nanobeam (using specifically placed notches as broadband couplers, see *Supplementary Information*), thus enabling the free-space transmission measurements. A set of representative transmission spectra, collected via super-continuum excitation and spectrometer, taken from the same device but at different input/collection polarizations, are shown in Figure 4 (c). These spectra correspond to TM-like (green curve) and TE-like (blue curve) polarized light transmitted through the diamond nanobeam waveguide which contains the optical cavity. Stop band (no transmission) and pass bands (high transmission) of the photonic crystal are clearly seen (the approximate location of the transition indicated with a dashed grey line), with the sharp resonances in the stop band corresponding to cavity modes. High resolution spectra of the fundamental TM-like and TE-like cavity modes are shown in Figures 4 (d) and (e), revealing waveguide coupled Q-factors of $Q_{TM} \sim 4,400$ and $Q_{TE} \sim 5,100$, respectively. Of the fabricated visible band nanobeam cavities, our best device had a measured Q-factor of $Q_{TE} \sim 8,200$.

In summary, high Q-factor racetrack and photonic crystal nanobeam cavities, realized using *angled-etching* nanofabrication scheme, have been demonstrated in bulk single-crystal diamond. Our devices feature Q-factors on par with those typically found in devices fabricated by conventional means, in standard photonic materials. Considering their wavelength scale mode volume ($V \sim (\lambda/n)^3$), photonic crystal nanobeam cavities shown here feature the highest *Q/V* figure of merit demonstrated in single-crystal diamond to date. We have also showed that low loss waveguides carved from a bulk diamond crystal can be made using our novel tapered vertical support structures over a broad wavelength range. Our results demonstrate that single-crystal diamond is a viable nanophotonics platform, and will enable further breakthroughs in both classical and quantum optics. For instance, when fabricated around spectrally stable NV⁻ or other color centers[16-17,40], diamond cavities will enable large enhancement of zero-phonon line emission via the Purcell effect, as well as efficient collection of emitted photons. We emphasize that monolithic single-crystal diamond nanophotonic structures are ultimately compatible with post processing techniques needed to stabilize implantation-defined color centers, which often include high temperature (~ 1200 °C) annealing[41]. Moreover, high-Q optical nanocavities in diamond are an attractive nonlinear optics platform[18], and would combine the advantage of relatively large Kerr nonlinearity and large Raman gain, lack of two- or multi-photon absorption, and excellent thermal properties for the generation of on-chip high repetition rate frequency combs[18,42] and Raman lasers at exotic wavelenths[43-44]. The free-standing nature of angled-etched nanophotonic devices also offers mechanical degrees of freedom, allowing exploration of diamond optomechanics[45], thus leveraging diamond's unique mechanical and optical properties. Combining diamond optomechanical devices with NV⁻ centers can result in on-chip hybrid quantum systems that rely on coherent spin-phonon-photon interactions for spin transduction and quantum state transfer[46-47]. Finally, our work will pave the way to realization of on-chip integrated photonic networks in other crystalline materials for which thin-film technology is not readily available.

This work was supported in part by the Defense Advanced Research Projects Agency (QuINESS program), and AFOSR MURI (grant FA9550-12-1-0025). Fabrication was performed at the Center for Nanoscale Systems (CNS) at Harvard University. M.J. Burek is supported in part by the Natural Science and Engineering Council (NSERC) of Canada and the Harvard Quantum Optics Center (HQOC). The authors thank A. Woolf and E. Hu for assistance with their visible supercontinuum source, and thank V. Venkataraman and H. Atikian for valuable discussions.

## METHODS

### Angled-etching nanofabrication

Standard optical grade, <100>-oriented, single-crystal diamond substrates (CVD grown, type IIa, < 1 ppm [N], Element Six) were polished to yield a surface roughness < 5 nm RMS (performed by Delaware Diamond Knives). Received polished diamond substrates were then cleaned in a boiling mixture consisting of equal parts sulfuric acid, nitric acid, and perchloric acid, followed by a pre-fabrication surface preparation[48] performed in a UNAXIS Shuttleline ICP-RIE. This included a 30 minute etch with the following parameters: 400 W ICP power, 250 RF power, 40 sccm Ar flow rate, 25 sccm $Cl_2$ flow rate, and 8 mTorr chamber pressure, followed by a second 30 minute etch with the following parameters: 700 W ICP power, 100 RF power, 50 sccm $O_2$ flow rate, and 10 mTorr chamber pressure. The purpose of this pre-fabrication step was to reduce the surface roughness to < 1 nm RMS and remove several microns from the top of the diamond substrate which is likely strained due to initial mechanical polishing.

Following surface preparation, a silica etch mask was patterned on the diamond substrates using hydrogen silsesquioxane (HSQ, FOX®-16 from Dow Corning) negative resist and electron beam lithography. Exposed HSQ was developed in tetramethylammonium hydroxide (TMAH, 25% diluted solution). The silica etch mask pattern was transferred into the diamond via a conventional top down anisotropic plasma etch – also in the UNAXIS Shuttleline ICP-RIE – with the following parameters: 700 W ICP power, 100 RF power, 50 sccm $O_2$ flow rate, 2 sccm $Cl_2$ flow rate, and 10 mTorr chamber pressure. The diamond was etched to a depth between 600 and 1000 nm, depending on the particular device being fabricated. Following this, the angled-etching step was performed to realize the free-standing nanophotonic devices. Angled-etching was achieved using the same ICP-RIE parameters as the initial top down etch, but included housing the sample inside a specifically designed aluminum Faraday cage[20] to direct the plasma ions to the substrate surface at the intended angle *(refer to Supplementary Information* for additional description). Two different Faraday cage designs were used (described in detail in *Supplementary Information*), one which was constructed to target a 45º etch angle, and the other for a target 60º etch angle. The resulting etch angles from these cage designs were measured to be approximately 35º and 50º, respectively. All diamond nanobeam cavities were fabricated at a 35º etch angle, while all diamond racetrack resonators were fabricated at a 50º etch angle. Following the oxygen-based plasma etching, the remaining etch mask was removed in concentrated hydrofluoric acid. Nanophotonic devices were cleaned in piranha solution prior to characterization.

**Fiber taper characterization at telecom and visible wavelengths**

Transmission measurements in the telecom band were collected via fiber taper coupling to suspended single-crystal diamond racetrack resonators. Fiber tapers were manufactured from SMF-28 fiber by the conventional flame anneal and pulling method[24], resulting in a final diameter of ~ 1 μm. The fiber taper was mounted in a U-shaped configuration, resulting in self-tension of the taper region and allowing it to be position in close proximity to the desired diamond device. Since the diamond nanophotonic devices are positioned above the diamond substrate in excess of 2 microns, dimpling the fiber taper was not necessary.

The fiber taper was spliced into an optical set-up, and its position with respect to the device under test was precisely controlled via motorized stages with 50 nm encoder resolution. Two tunable lasers (Santec TSL-510, tuning range from 1480 to 1680 nm) were used, along with an inline fiber polarizer, and high gain InGaAs detector (EO Systems, IGA1.9-010-H) to record transmission spectra. All collected spectra were normalized by transmission data collected from an uncoupled position.

For transmission measurements at visible wavelengths, fiber tapers were manufactured from commercial SM-600 fiber via wet etching in hydrofluoric acid[49]. Bare SM-600 fiber was again mounted in a U-shape configuration, followed by localized wet etching in hydrofluoric acid near the center. The hydrofluoric acid was covered with a thin layer of o-xylene on top in order to promote gradual taper formation via the oil/water interface meniscus. A two step etch process which included ~ 30 minutes of etching in concentrated hydrofluoric acid, followed by ~ 30 to 50 minutes etching in 5:1 buffered oxide etch (BOE) was used to thin the final taper region to a diameter of ~ 500 to 700 nm. Following visible fiber taper formation, the mounted fiber was again spliced into the same physical set up as described for telecom band measurements. A fiber coupled supercontinuum laser source (EXW-4, NKT Photonics) and optical spectrum analyzer (OSA, HP 70950B, minimum resolution bandwidth of 0.08 nm) were used to collect broadband spectra in roughly the 680 to 800 nm band. To gauge the resolution of the OSA, a HeNe laser was connected directly to the OSA. Measurement of the resulting laser emission spectra yielded a line width ~ 70 pm, which is artificially broadened by the instrument resolution. As such, any measured Q-factor near and above 9,000 was deemed resolution limited. To avoid this resolution limitation, a tunable red laser (New Focus Velocity TLB 6304 laser, coarse tuning range of 634.8 to 638.9 nm and fine tuning range of 70 pm) and visible band photodetector (New Focus 1801) were also used to collect transmission spectra. The fine tuning range of the laser was used to accurately measure the Q-factors of cavity modes supported by the diamond racetrack resonators within the ~ 4 nm coarse laser tuning range.

**Free space transmission measurement at visible wavelengths**

Transmission measurements conducted by free-space coupling utilized a home built confocal microscope in which a high numerical aperture (NA = 0.95) objective was used to focus laser light onto the sample. The input laser optical path was scanned using a galvo mirror imaged onto the back of the objective with a pair of lenses comprising a 4*f* imaging system. Additionally, a beam splitter was placed between the input channel galvo mirror and the 4*f* imaging system in order to incorporate a collection channel with independent scanning control via its own set of galvo mirrors. The free-space coupling set up was thus able to pump and collect light at two spatially separated positions, which allowed for free-space transmission measurements. Positioning of a target device under the objective was accomplished by precision motorized stages. In order to couple light into and out of the diamond nanobeams, local broadband scatter centers in the form of notches – as seen in Figure 1 (d) of the main text – were incorporated at the two ends of fabricated nanobeam cavities. Light from a supercontinuum laser source (EXW-4, NKT Photonics) was coupled into the diamond nanobeam cavity, with light out-coupled from the structure sent to a spectrometer. Free-space polarizers were also included in both the input and collection channels. A schematic of the free-space transmission measurement optical setup is included in the *Supplementary Information*.

**FIGURE CAPTIONS**

**Figure 1 | Angled-etching fabrication methodology. (a)** Illustration of angled-etching used to realize free-standing structures in bulk single-crystal diamond. **(b)** Angled-etching fabrication steps: (i) define an etch mask on substrate via standard fabrication techniques, (ii) transfer etch mask pattern into the substrate by conventional top down plasma etching, (iii) employ angled-etching to realize suspended nanobeam structures, (iv) remove residual etch mask. SEM images of **(c)** a fabricated diamond racetrack resonator supported from the bottom and **(d)** a fabricated diamond nanobeam photonic crystal cavity operating at visible wavelengths. All SEM images were taken at a $60^o$ stage tilt.

**Figure 2 | High-Q diamond racetrack resonators.** SEM images of **(a)** 25 μm bend radius diamond racetrack resonator, with close-up **(b)** side and **(c)** top views. The nominal ($w_o$) and maximum ($w_s$) width (indicated in figure) of the tapered vertical are approximately 1.1 μm and 1.27 μm, respectively. Note, a ~ $50^o$ etch angle was used to fabricate devices shown. All SEM images were taken at a $60^o$ stage tilt. Simulated **(d)** TE-like and **(e)** TM-like mode profiles ($\lambda$ = 1.55 μm, electric field norm) of a suspended 1.1 μm wide diamond waveguide. **(f)** Representative normalized broadband spectrum of a 1.1 μm wide and 37.5 μm bend radius diamond racetrack resonator collected by tapered fiber measurement. Inset shows optical micrograph indicating the tapered fiber coupling position. High-resolution spectra of near critically coupled **(g)** TE-like and **(h)** TM-like modes with loaded Q-factors indicated.

**Figure 3 | High-Q diamond nanobeam photonic crystal cavities.** SEM images of **(a)** diamond photonic crystal nanobeam cavities, with close-up **(b)** prospective and **(c)** top down views. Note, a ~ $35^o$ etch angle was used to fabricate devices shown. All SEM images were taken at a $60^o$ stage tilt.

Simulated cross-sectional and top down electric field intensity profiles of the fundamental **(d)** TE-like and **(e)** TM-like photonic crystal cavity modes. Note, top down mode profiles correspond to top face of the nanobeam cavity. **(f)** Representative normalized broadband spectrum of a fabricated diamond photonic crystal cavity collected by tapered fiber measurement, with inset optical micrograph indicating the tapered fiber coupling position. High resolution spectra of the fundamental **(g)** TM-like and **(h)** TE-like cavity modes. The taper-loaded Q-factors of the fundamental and second order TM-like cavity modes were 24,000 and 3,700 respectively, while the first three TE-like cavity modes had loaded Q-factors of 183,000, 94,000, and 22,000, respectively.

**Figure 4 | Diamond optical nanocavities at visible wavelengths.** **(a)** Normalized broadband transmission spectrum collected by fiber taper coupling from a diamond racetrack resonator (17.5 μm bend radius and ~ 500 nm beam width) using: i) a tunable red laser and photodiode in 635 nm to 639 nm range (grey curve), and ii) super-continuum source and spectrometer in 688 nm to 720 nm range (red curve). Two sets of supported resonances – the TE-like and TM-like waveguide modes – are again apparent. The insets reveal the fiber taper coupling position with a red laser tuned off and on resonance with the optical cavity. **(b)** Corresponding high resolution spectra of two cavity modes near ~ 637 nm (loaded Q-factor indicated on plot) collected via a tunable laser. **(c)** Representative broadband transmission spectra of a fabricated diamond nanobeam photonic crystal cavity operating in visible collected via free space coupling at different input polarizations. High resolution spectra corresponding to the fundamental **(d)** TM-like and **(e)** TE-like cavity modes, revealing waveguide coupled Q-factors of 4,400 and 5,100, respectively.

**FIGURES**

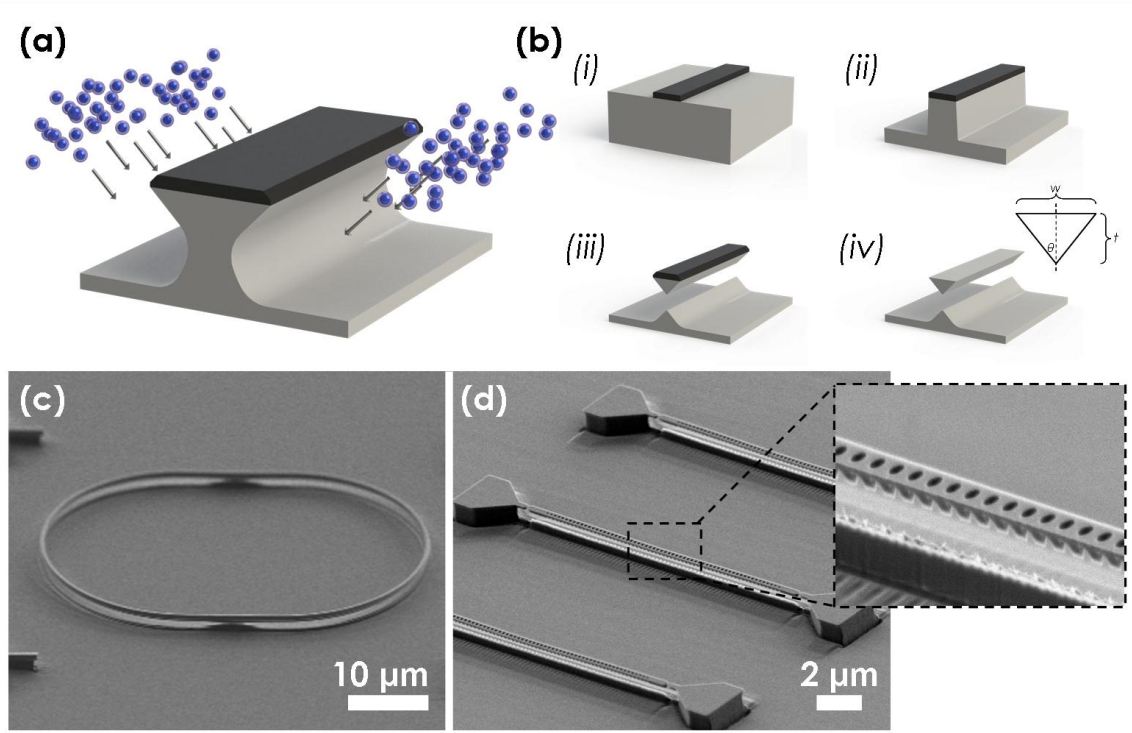

**Figure 1 | Angled-etching fabrication methodology. (a)** Illustration of angled-etching used to realize free-standing structures in bulk single-crystal diamond. **(b)** Angled-etching fabrication steps: (i) define an etch mask on substrate via standard fabrication techniques, (ii) transfer etch mask pattern into the substrate by conventional top down plasma etching, (iii) employ angled-etching to realize suspended nanobeam structures, (iv) remove residual etch mask. SEM images of **(c)** a fabricated diamond racetrack resonator supported from the bottom and **(d)** a fabricated diamond nanobeam photonic crystal cavity operating at visible wavelengths. All SEM images were taken at a 60° stage tilt.

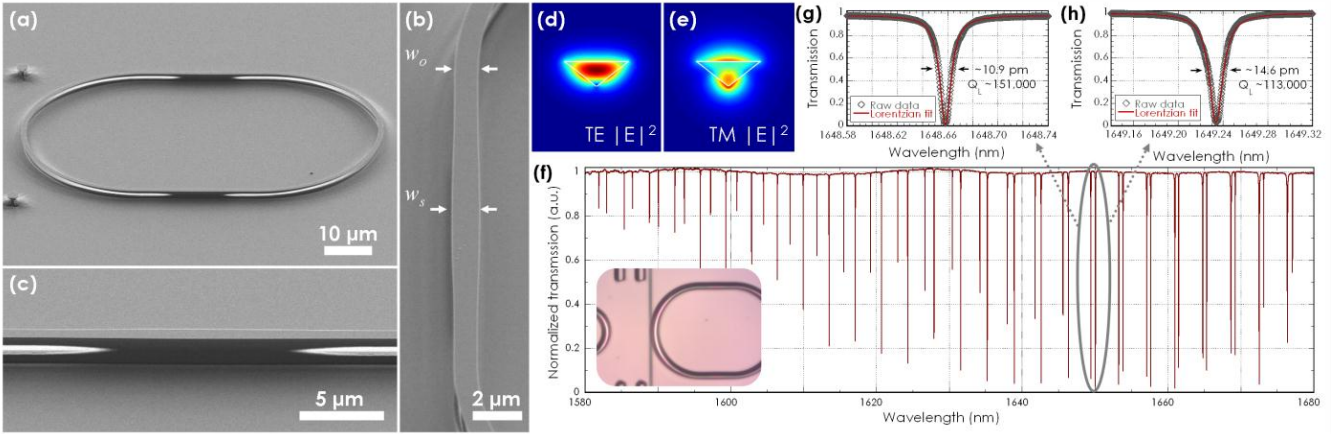

**Figure 2 | High-Q diamond racetrack resonators.** SEM images of **(a)** 25 μm bend radius diamond racetrack resonator, with close-up **(b)** side and **(c)** top views. The nominal ($w_o$) and maximum ($w_s$) width (indicated in figure) of the tapered vertical are approximately 1.1 μm and 1.27 μm, respectively. Note, a ~ 50° etch angle was used to fabricate devices shown. All SEM images were taken at a 60° stage tilt. Simulated **(d)** TE-like and **(e)** TM-like mode profiles ($\lambda$ = 1.55 μm, electric field norm) of the suspended 1.1 μm wide diamond waveguide. **(f)** Representative normalized broadband spectrum of a 1.1 μm wide and 37.5 μm bend radius diamond racetrack resonator collected by tapered fiber measurement. Inset shows optical micrograph indicating the tapered fiber coupling position. High-resolution spectra of near critically coupled **(g)** TE-like and **(h)** TM-like modes with loaded Q-factors indicated.

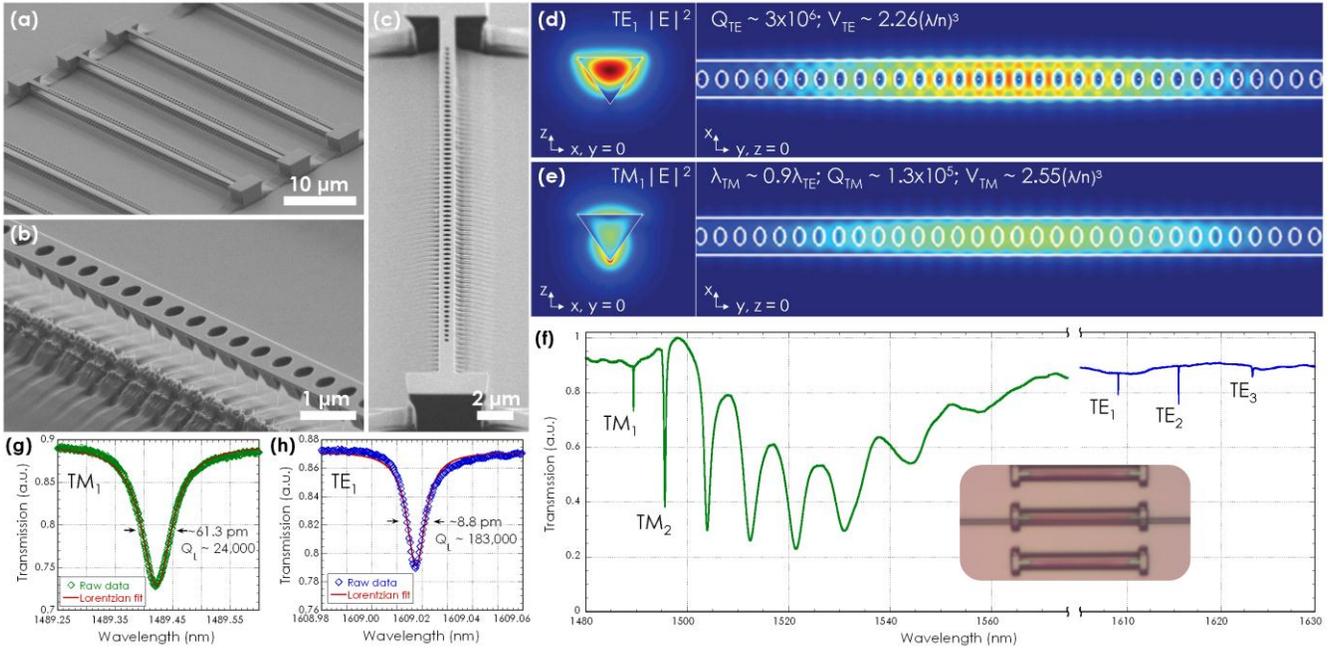

**Figure 3 | High-Q diamond nanobeam photonic crystal cavities.** SEM images of **(a)** diamond photonic crystal nanobeam cavities, with close-up **(b)** prospective and **(c)** top down views. Note, a ~ 35° etch angle was used to fabricate devices shown. All SEM images were taken at a 60° stage tilt. Simulated cross-sectional and top down electric field intensity profiles of the fundamental **(d)** TE-like and **(e)** TM-like photonic crystal cavity modes. Note, top down mode profiles correspond to top face of the nanobeam cavity. **(f)** Representative normalized broadband spectrum of a fabricated diamond photonic crystal cavity collected by tapered fiber measurement, with inset optical micrograph indicating the tapered fiber coupling position. High resolution spectra of the fundamental **(g)** TM-like and **(h)** TE-like cavity modes. The taper-loaded Q-factors of the fundamental and second order TM-like cavity modes were 24,000 and 3,700 respectively, while the first three TE-like cavity modes had loaded Q-factors of 183,000, 94,000, and 22,000, respectively.

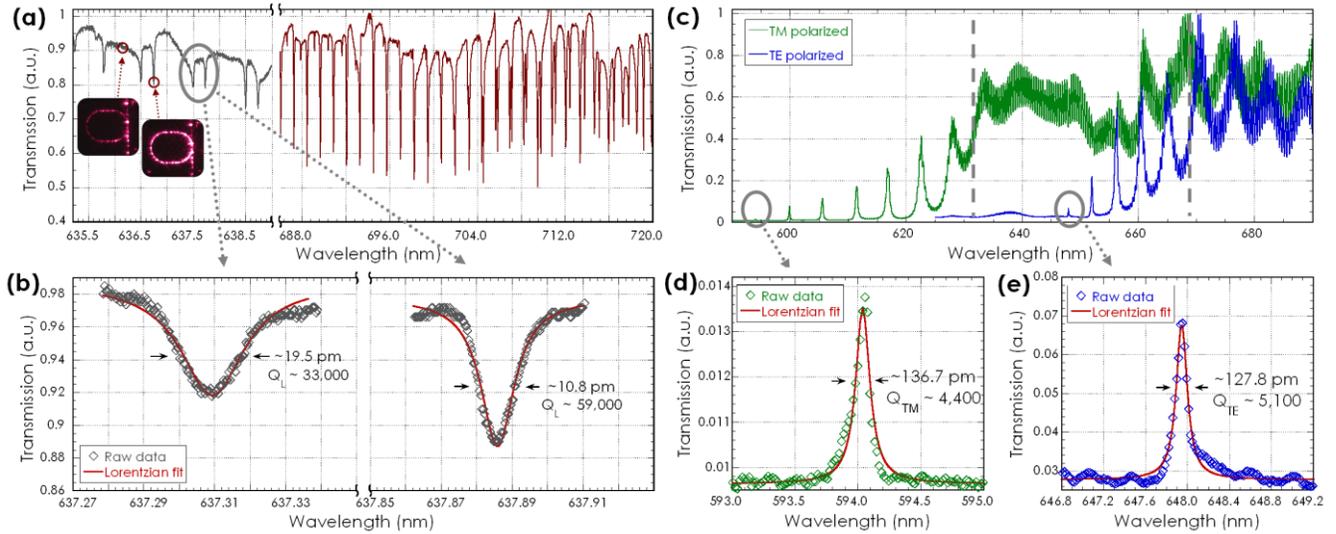

**Figure 4 | Diamond optical nanocavities at visible wavelengths. (a)** Normalized broadband transmission spectrum collected by fiber taper coupling from a diamond racetrack resonator (17.5 μm bend radius and ~ 500 nm beam width) using: i) a tunable red laser and photodiode in 635 nm to 639 nm range (grey curve), and ii) super-continuum source and spectrometer in 688 nm to 720 nm range (red curve). Two sets of supported resonances − the TE-like and TM-like waveguide modes − are again apparent. The insets reveal the fiber taper coupling position with a red laser tuned off and on resonance with the optical cavity. **(b)** Corresponding high resolution spectra of two cavity modes near ~ 637 nm (loaded Q-factor indicated on plot) collected via a tunable laser. **(c)** Representative broadband transmission spectra of a fabricated diamond nanobeam photonic crystal cavity operating in visible collected via free space coupling at different input polarizations. High resolution spectra corresponding to the fundamental **(d)** TM-like and **(e)** TE-like cavity modes, revealing waveguide coupled Q-factors of 4,400 and 5,100, respectively.

# SUPPLEMENTARY INFORMATION

### i) Angled-etching of free-standing diamond nanobeam structures

To supplement the nanofabrication details given in the *Methods* section of the main text, plasma etching of diamond nanostructures is performed in a UNAXIS Shuttleline inductively coupled plasma-reactive ion etcher (ICP-RIE) with the following parameters: 700 W ICP power, 100 RF power, 50 sccm $O_2$ flow rate, and 10 mTorr chamber pressure. These ICP-RIE parameters are used for conventional anisotropic etching of diamond structures at a rate of approximately 200 nm/min, and were optimized for smooth, near vertical side walls[1-2]. Using these parameters, angled-etching is achieved by housing the diamond substrate within a specifically designed aluminium Faraday cage, which is subsequently placed within this ICP-RIE system. The Faraday cage shields the diamond substrate from the electromagnetic fields which build up inside the ICP-RIE system and are responsible for acceleration of plasma ions towards the sample surface. Although the Faraday cage has small grid openings on its surface, the effect of an external field is attenuated drastically within a small distance of the openings. Thus, during the plasma etching process, the potential gradient builds up over the face of the Faraday cage and accelerates ions along a path perpendicular to the cage surface. After ions move through the potential gradient and past the metal grid, they are no longer accelerated and travel virtually unimpeded inside the cage towards the substrate. Therefore, with a properly designed Faraday cage geometry, plasma ions may be directed to the sample surface at an oblique angle in multiple directions[3]. It is important to emphasize that while different configurations of Faraday cages can be used, angled-etching is not realized through simple tilting of the substrate within a plasma etcher without a Faraday cage[4,5].

In this work, Faraday cages consisted of two parts: (1) a structural base machined from aluminum and

(2) commercially available wire mesh (also aluminium). The mesh was woven 250 μm diameter wire at a 2 mm x 2 mm pitch. The meshed was bent around the machined structural base and fixed with aluminium bolts, thus forming the completed Faraday cage. Aluminium was used since it is not chemically attacked and does not erode at a significant rate during the plasma etching process, and forms only a thin, stable oxide layer.

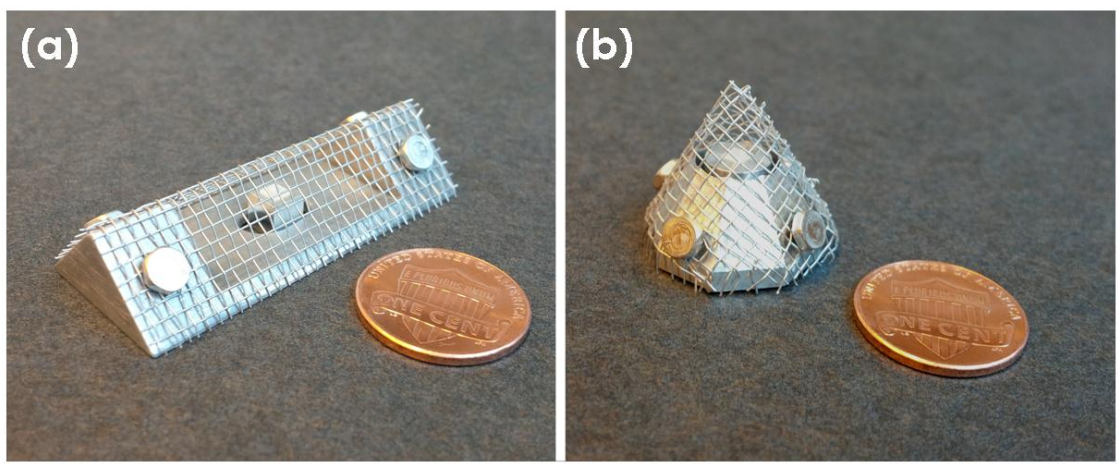

**Figure S1.** Images of constructed Faraday cage designed used in this work: **(a)** triangular prism Faraday cage with incline angle of $\theta = 45°$, and **(b)** conical Faraday cage with incline angle of $\theta = 60°$.

Two specific Faraday cage designs were used in this work. The main design constraint for the Faraday cages was the load-lock height clearance of our ICP-RIE system, which was approximately 20 mm. The first Faraday cage design is a triangular prism structure shown in Figure S1 (a), which allowed for angled-etching in two simultaneous directions[1]. The triangular prism Faraday cage in Figure S1 (a) was used for the fabrication of diamond nanobeam cavities described in the main text, operating at both telecom and visible wavelengths. The ion incidence angle relative to the substrate surface normal (the etch angle, $\theta$) is defined by incline angled of the Faraday cage. The incline angle of this Faraday cage was $\theta = 45°$, and its height and length were 10 mm and 50 mm respectively. The distance between cage bottom and the surface of the diamond substrate fixed at ~ 3.5 mm using an aluminium sample mounting block. A hole in the bottom of the cage allowed it to be placed over the mounted sample on the ICP-RIE wafer carrier.

The second Faraday cage used in this work was a conical design shown in Figure S1 (b), where angled-etching occurred in all directions simultaneously[1]. A conical Faraday cage design enables the fabrication of free-standing diamond nanobeams with arbitrary curvatures, and thus, was used to produce diamond racetrack resonators described in the main text operating at both telecom and visible wavelengths. The conical Faraday cage was constructed in a similar fashion as the triangular prism Faraday cage, and had a bottom diameter of 20 mm with an incline angle of $\theta = 60°$. The diamond substrate surface fixed at a height of ~ 10 mm from the cage bottom in a similar fashion as before.

**ii) Cross-sectional analysis of diamond nanobeams fabricated by angled-etching**

In our nanofabrication scheme, the etch angle is dictated by the physical design of the Faraday cage used to accomplish angled-etching[1]. In this work, two different Faraday cages were used (as described in detail in the previous section), one which was constructed to target a 45° etch angle, and the other for a target 60° etch angle. Focused ion beam (FIB) milled cross-sections were inspected to measure the approximate etch angle resulting from the two cage designs. To properly cross-section free-standing diamond nanobeams by FIB milling, the diamond substrate with free-standing nanobeams was first sputter coated with a ~ 5 nm titanium adhesion layer, followed by ~ 150 nm of gold. The initial gold coating was necessary to avoid charging of the electron and ion beams in the dual beam SEM-FIB system (Zeiss Nvision 40). Prior to FIB milling, several selected diamond nanobeams were further coated in a thick (several micron) coating of platinum by localized ion beam assisted deposition. The platinum coating is necessary to achieve a clean cross-section and avoid rounded edges due to inadvertent material erosion from beam tails and ion scattering. After the localized platinum deposition, free-standing nanobeams were cross-sectioned by rastering the ion beam in a rectangular pattern which was aligned perpendicular with the nanobeam axis. The result was a relatively clean cross-section of the diamond nanobeam which was subsequently inspected in SEM. Examples are shown in Figure S2, with the gold and platinum coatings clearly visible.

Figure S2 shows representative FIB milled cross-sections of diamond nanobeams fabricated with each of the two Faraday cages designs. The corresponding etch angles measured to be ~ 35° and 50°, respectively. The 10° deviation from each target etch angle may be attributed to several factors, including the slight misalignment of the FIB milled cross-section from the nanobeam axis, the non-ideal verticality of the $O_2$ plasma etch conditions for single-crystal diamond, and also a dependence of etch angle on sample height observed within the Faraday cage (i.e. samples near the bottom of the cage etch at slightly steeper angles due to the plasma parameters and small Faraday cage size). From the cross-sections, the diamond nanobeams appear fairly symmetric, with only a slight degree of edge roughness. However, it is not clear whether this roughness is the result of the FIB milling process or the angled-etching process.

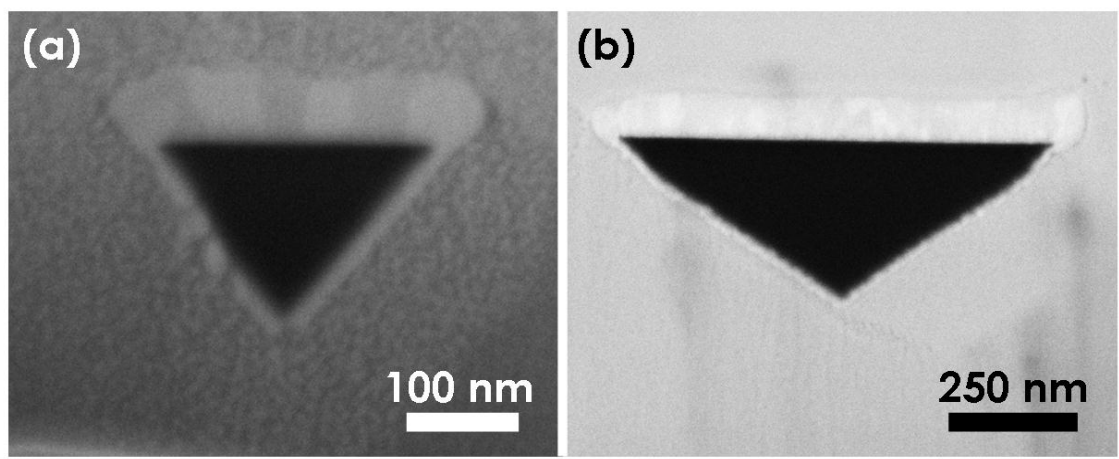

**Figure S2.** SEM images of FIB cross-sectioned diamond nanobeams fabricated with etch angles of **(a)** ~ 35°, and **(b)** ~ 50°. All SEM images were taken at a 60° stage tilt.

While the FIB milled cross-sections give an estimate of the etch angle, it is important to note that the difficulty and serial nature of FIB milling make it impossible to sample a large number of angled-etched diamond nanobeams. To truly evaluate the etch angle of an ensemble of diamond nanobeams, the etch angle may be estimated by evaluating low frequency flexural mechanical resonances of free-standing diamond nanobeam cantilevers, as was previously reported by us[6]. Briefly – from Euler-Bernoulli elastic beam theory – the ratio of the fundamental out-of-plane and in-plane flexural mechanical resonance frequencies reflects the etch angle through the relation: $\theta = \tan^{-1}\left(f_y / \sqrt{3} f_x\right)$, where $f_x$ and $f_y$ are the out-of-plane and in-plane mechanical resonance frequencies. The mechanical resonance frequencies of a series of free-standing diamond nanobeam cantilevers fabricated with both the triangular prism and conical Faraday cages were measured, with over 100 cantilevers sampled for each cage design. Applying the above relation to the experimental data gave an etch angle of $33.9° \pm 1.4°$ and $53.5° \pm 1.5°$ for the triangular prism and conical Faraday cages, respectively. These values are truly ensemble estimates and in close agreement with those obtained from FIB milled cross-sections.

To further evaluate the surface quality of the angled-etching diamond nanobeams, several angled-etched nanobeams were physically removed using a tungsten probe controlled by micromanipulators, and transferred to a silicon substrate for subsequent SEM inspection. The transferred beams were approximately 400 nm wide and angled-etch under the same conditions used to fabricate nanobeam photonic crystal cavities similar to those shown in Figure 1 (d) of the main text. Figure S3 shows representative images of two transferred diamond nanobeams, with the nanobeam in Figure S3 (a) oriented upside down revealing two angled-etched surfaces, and the nanobeam displayed in Figure S3 (b) oriented on its side revealing the top surface and one angled etched surface. From these SEM images, comparable surface roughness is observed between the angled-etch surfaces and the top diamond surface, with the exception of slight roughness localized to the just beneath the top corners of the diamond nanobeams. This concentration of surface roughness likely originates from the erosion of the etch mask during angled-etching. In general, the angled-etched diamond nanobeams inspected in this manner were smooth and fairly symmetric.

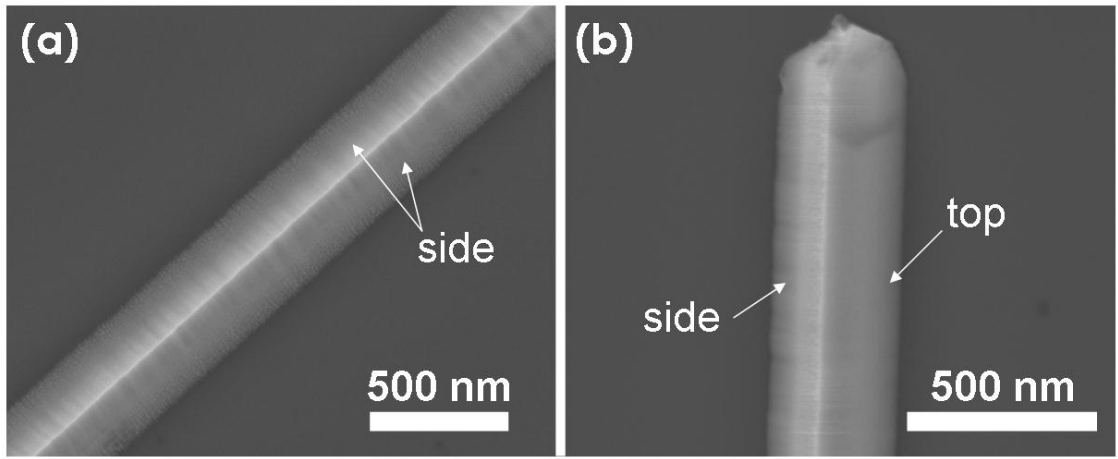

**Figure S3.** SEM images angled-etched ~ 400 nm wider diamond nanobeams which have been physically removed via a micromanipulated tungsten probe and transferred to a silicon substrate. The transferred diamond nanobeams are oriented **(a)** upside down revealing two angled-etched surfaces and **(b)** on its side revealing the top surface and one angled-etched surface.

**iii) Modal analysis of vertically support diamond waveguides**

To confirm a vertically supported, triangular cross-section diamond nanobeam does function as a low loss optical waveguide, eigenmode solver software (MODE Solutions, Lumerical) was used to simulate the parameters of the ideal vertical support cross-section. Figure S4 (a) and (b) show fundamental TE-like and TM-like mode profiles of the support cross-section, respectively, for a propagation wavelength of $\lambda$ = 1.55 µm. Here, the nanobeam width is increased 15 % from a nominal waveguide width of 1.1 µm, resulting in a 165 nm wide diamond "fin" supporting the waveguide. The distance between the diamond substrate and the bottom of the diamond waveguide was varied from 1 to 2 microns. Key parameters of the simulation were a mesh size of 30 nm and a 6 x 6 µm$^2$ perfectly matched layer (PML).

The optical loss estimated from simulation, which is plotted in Figure S4 (c), confirmed the ideal support cross-section can be well below 1 dB/cm for both supported modes, given the separation between the bottom of the waveguide and substrate exceeds ~ 1.8 μm. Additionally, increasing the relative width modulation to 25%, while keeping the width of diamond fin at 165 nm, lowers the optical loss slightly from the nominal design.

Figure S4 (d) shows a FIB milled cross-section of the pedestal-like structure achieved in the vicinity of the maximum width of the vertical support. Note this cross-section is taken from the diamond racetrack resonator which produced the data shown in Figure 2 of the main text. The separation between diamond waveguide and the substrate for this structure is nearly 2 μm. While not detrimental to device performance (as confirmed experimentally), irregularity in the pedestal-like cross-section is clearly observed and attributed to an etch rate anisotropy, where material was removed at a slightly faster rate from the right side of the support shown in Figure S4 (d).

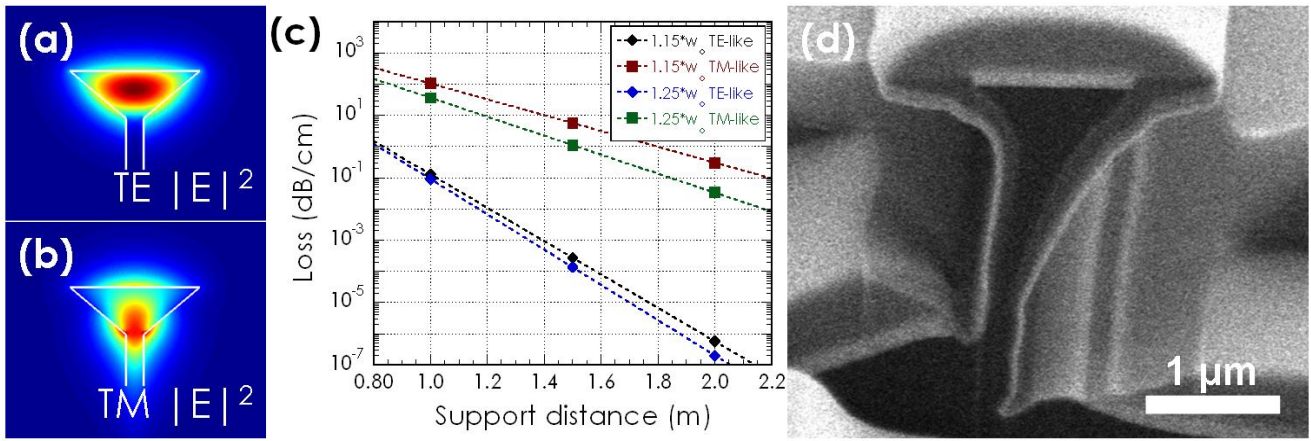

**Figure S4.** Simulated transverse mode profiles ($\lambda$ = 1.55 μm, electric field norm) of the 1.265 μm wide **(a)** TE-like and **(b)** TM-like vertical support waveguide modes. **(c)** Plotted optical loss estimated from simulations for the TE/TM-like modes of a diamond vertical support structure as a function of separation between the bottom of the diamond waveguide and substrate (labeled 'Support distance'). **(d)** SEM image of FIB cross-sectioned tapered vertical support near its maximum width. All SEM images were taken at a 60° stage tilt.

### iv) Free-space transmission measurement set up

As described in the *Methods* section of the main text, a schematic of the free-space transmission measurement optical setup is displayed in Figure S5 (a). An illustration indicating how light is in- and out-coupled at opposite ends of the nanobeam (using specifically placed notches as broadband couplers) is shown in Figure S5 (b). The coupling efficiency of notches in the diamond nanobeam was not fully characterized as it is outside the scope of this work.

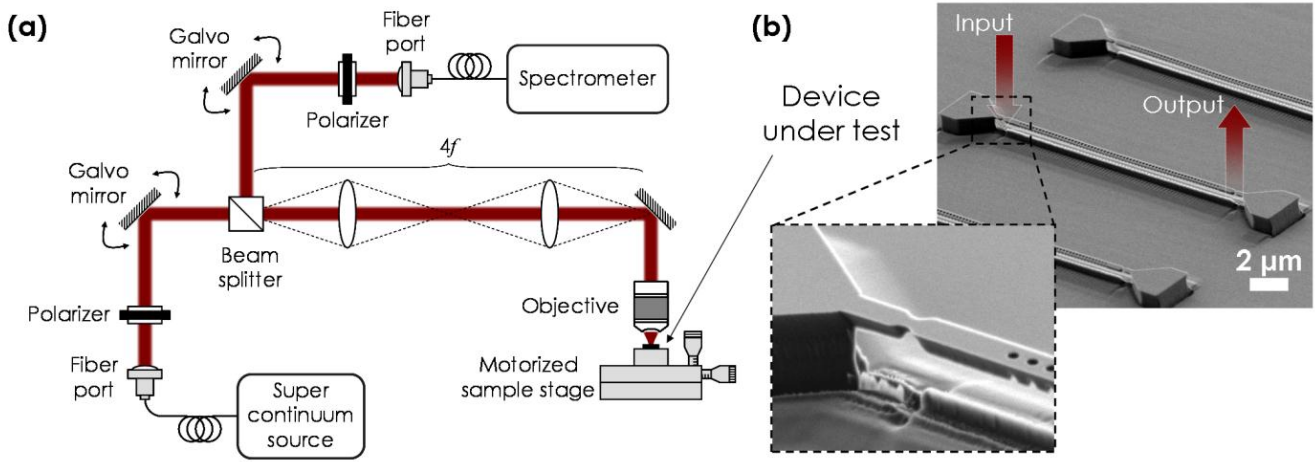

**Figure S5. (a)** Schematic of the free-space transmission measurement setup used to characterize diamond nanobeam cavities operating at visible wavelengths. See the *Methods* section of the main text for a detailed description. **(b)** SEM image of a diamond nanobeam cavity (same as Figure 1 (a) of the main text) with a zoomed in image of a specifically placed notch in the diamond nanobeam included as an inset. The locations where light is in- and out-coupled from the diamond nanobeam cavity via notches are illustrated on the image.